\documentclass[12pt]{article}

\usepackage{theorem,amssymb,amsbsy,latexsym}

\newcommand{\nd}{{\vphantom{\dag}}}

\newcommand{\cH}{{\cal H}}
\newcommand{\cF}{{\cal F}}
\newcommand{\cE}{{\cal E}}
\newcommand{\cZ}{{\cal Z}}
\newcommand{\cN}{{\cal N}}
\newcommand{\ccS}{{\cal S}}
\newcommand{\tcS}{\tilde{\cal S}}

\newcommand{\bcE}{\cE^0}
\newcommand{\cc}{r}
\newcommand{\gl}{{\rm gl}}
\newcommand{\vv}{c}

\newcommand{\vl}{{\bf l}}
\newcommand{\vm}{{\bf m}}
\newcommand{\vx}{{\bf x}}
\newcommand{\vy}{{\bf y}}
\newcommand{\vk}{{\bf k}}
\newcommand{\vp}{{\bf p}}
\newcommand{\vq}{{\bf q}}
\newcommand{\Ref}[1]{(\ref{#1})}
\newcommand{\Tr}{{\rm Tr}}

\newcommand{\R}{{\mathbb R}}
\newcommand{\C}{{\mathbb C}}

\newcommand{\eq}{\begin{equation}}
\newcommand{\eqend}{\end{equation}}
\newcommand{\eqa}{\begin{eqnarray}}
\newcommand{\nonueqa}{\begin{eqnarray*}}
\newcommand{\eqaend}{\end{eqnarray}}
\newcommand{\nonueqaend}{\end{eqnarray*}}
\newcommand{\nonu}{\nonumber \\ \nopagebreak}
\newcommand{\bma}[1]{\begin{array}{#1}}
\newcommand{\ema}{\end{array}}
\newcommand{\bc}{\begin{center}}
\newcommand{\ec}{\end{center}}

\begin{document}
\begin{flushright}
\date{May 28, 2002} 
\end{flushright}
\vspace{.4cm}
\begin{center}

{\Large Interacting fermions on noncommutative spaces:
Exactly solvable quantum field theories in $2n+1$ dimensions:}\\

\vspace{1 cm}

{\large Edwin Langmann}\\

\vspace{0.3 cm}
{\it Mathematical Physics, Royal Institute of Technology, SE-10691 Stockholm, Sweden}

{\tt langmann@theophys.kth.se}

\end{center}

\newcommand{\fintf}[1]{\int \frac{d^4 #1}{(2\pi)^4}}
\newcommand{\fintt}[1]{\int \frac{d^2 #1}{(2\pi)^2}}

\begin{abstract}

I present a novel class of exactly solvable quantum field
theories. They describe non-relativistic fermions on even dimensional
flat space, coupled to a constant external magnetic field and a four
point interaction defined with the Groenewold-Moyal star product.
Using Hamiltonian quantization and a suitable regularization, I show
that these models have a dynamical symmetry corresponding to
$\gl_\infty\oplus \gl_\infty$ at the special points where the magnetic
field $B$ is related to the matrix $\theta$ defining the star product
as $B\theta=\pm I$. I construct all eigenvalues and eigenstates of the
many-body Hamiltonian at these special points. I argue that this
solution cannot be obtained by any mean-field theory, i.e.\ the models
describe correlated fermions.  I also mention other possible
interpretations of these models in solid state physics.

\end{abstract} 

\noindent {\bf 1. Introduction.} Exactly solvable models play a
special role in quantum field theory.  They provide a testing ground
for general methods and concepts applicable to other, more realistic,
models which can be studied only in approximations. Moreover, such
approximations usually are based on some exact solution.  The majority
of the known exactly solvable quantum field theories are free
(non-interacting), and most of the few known interacting ones are in
low dimensions and/or have supersymmetry (see e.g.\
\cite{Mattis,Lerche}).

In this paper I find and study a particular class of interacting field
theories without supersymmetry (as far as I can see) and which are
exactly solvable in $2n+1$ dimensional spacetime with $n=1,2,\ldots$
arbitrary. They describe non-relativistic fermions coupled to a
constant, external magnetic field and four-point interactions defined
with the Groenewold-Moyal star product.\footnote{To avoid
misunderstanding I stress that the free part of the model is the
standard Hamiltonian of fermions in an external magnetic fields: No
star product is used there.}  These models are specified by two
antisymmetric $2n\times 2n$ matrices $B$ and $\theta$ which define the
magnetic field and the star product, respectively (precise definitions
will be given below). Using the Hamiltonian framework, I will solve
these models for the special cases $B\theta=\pm I$ and arbitrary
interaction constant. A crucial point in the argument is to expand the
model in a convenient basis which brings it to a matrix form which
makes manifest a huge dynamical symmetry. It is interesting to note
that, in this latter form, this model also has a natural solid-state
physics interpretations: it also describes correlated fermions in
$2+1$ dimensions \cite{EL}.  Readers interested only in this aspect of
this work may skip all the discussion on non-commutative spaces and
start immediately with Section~4 below. Another aspect is the
mathematical structure involved in the solution: The above-mentioned
dynamical symmetry corresponds to the Lie algebra $\gl_\infty \oplus
\gl_\infty$, and, from a mathematical point of view, the solution
boils down to decomposing the regular representations of all
permutation groups into irreps.  The latter is a classical problem in
group theory, and I can solve the model by using some beautiful group
theory results obtained a long time ago (books containing proofs of
the results needed are, e.g.\ \cite{Mu,Ro,W}). However, I will first
derive the solution by an explicit construction exploiting the
dynamical symmetry. That method is perhaps less elegant but is, as I
hope, a good starting point for getting an intuitive understanding of
the solution. (The former elegant method is, to my opinion, only
easier for experts in group theory. One could also read the two
methods together as a pedestrian proof of certain group theory
results.)  The group theory point of view suggests further
generalizations of the model with additional (peculiar) $p$-body
interactions where $p= 3,4,\ldots$. These additional interactions
correspond to higher order Casimir operators and leave the model
solvable.

As mentioned, the models considered can be naturally interpreted as
field theories of fermions on a noncommutative phase space (time
remains commutative): using the star product implies that the components of
the spatial variables $\vx=(x^1,\ldots,x^{2n})$ are made
noncommutative
\eq
\label{star}
x^\mu\star x^\nu - x^\nu\star x^\mu = -2i\theta^{\mu\nu} 
\eqend
($\mu,\nu=1,\ldots,2n$).  Thus, generalizing a standard field theory
model by replacing the pointwise product of fields by the star product
can be interpreted as making space noncommutative. In a similar
manner, coupling the fermions to a magnetic field can be interpreted
as a further generalization where momentum space is made
noncommutative as well: one replaces the commutative momentum
operators by noncommutative ones, $-i\partial_\mu \to -i\partial_\mu
-B_{\mu\nu} x^\nu$. Field theory models with noncommutative space have
received much attention recently (for review and a fairly exhaustive
list of references I refer to \cite{KS,DN,S}). In particular, these
standard noncommutative field theories are known to simplify, become
soluble, for the limiting case $\theta=\infty$ \cite{GMS,MRW}. I show
here that there are other solvable cases in the extended family of
field theories with noncommutative phase spaces, namely all those
where $B\theta=\pm I$. It is interesting to note that $\theta=\infty$
is precisely the limit where the kinetic energy term is irrelevant,
whereas the solutions obtained in this paper allow to study the effect
of (a particular kind of) kinetic energy as well.

It is worth noting that the field theories considered here are special
in that they look the same in position and in Fourier space and, in
particular, for $B \theta =\pm I$ they are (essentially) invariant
under Fourier transformation \cite{LS}.  I believe that it is the
latter property which makes this model exactly soluble: the boson
models discussed in \cite{LS} seem to be soluble at these special
points as well \cite{LSZ}.  I also mention that there is a (classical)
field theory model of bosons which somewhat similar to the one we
solve and whose integrability was noted already in~\cite{Hoppe}. I
believe that the quantum version of this model should be solvable by
similar methods as the ones used here.

It is interesting to note that these field theories in $2+1$
dimensions may provide interesting models for a quantum Hall system: I
add to the standard free quantum Hall Hamiltonian a particular
four-point interaction of the fermions, and the resulting model is
exactly solvable. In fact, the model I solve is somewhat more general
in that there is also a term describing a confining electrical
potential.  I will elaborate this interpretation slightly in the
conclusions, but I should stress already here that my results (i.e.\
eigenvalues and eigenfunctions) provide only a first step to
understand the physics of this model.

The plan of the rest of this paper is as follows. In the next section
I first define the models in the context of noncommutative field
theories (Section~2). In Sections~3--6 I solve these models in 2+1
dimension. The key is two mathematical facts about Landau
eigenfunctions which allow to write the Hamiltonian in a matrix form
(Section~3). In Section~4 I give a precise meaning to the model using
a natural regularization and constructing its Hilbert space
representation. In this Section I also demonstrate the dynamical
symmetry of the model. I then give two methods of solution, one using
a somewhat pedestrian approach (Section~5), and another using group
theory (Section~6). The generalization of this to $2n+1$ dimensions in
Section~7 is remarkably simple: from an abstract point of view,
changing the dimension does not make much of a difference.  Section~8
contains a short discussion of generalized models with additional
$p$-body interactions ($p=3,4\ldots$).  I end in Section~9 with a few
remarks on possible applications, generalizations, and open
questions. For the convenience of the reader two appendices are added:
Appendix~A contains an elementary proof of the above-mentioned facts,
and Appendix~B is to exemplify the general group theory results.

\bigskip

\noindent {\bf 2. Definition of the models.}  The models we consider
are defined by a Hamiltonian $\cH=\cH_0 + \cH_{\rm int}$ where the
free part is
$\cH_0 = \int d^{2n} \vx\, \Psi^\dag(\vx) H_B
\Psi(\vx)$ 
with 
\eq
\label{HL}
H_{B} = (-i\partial_\mu -  B_{\mu\nu}x^\nu)^2 
\eqend 
the (generalized one-particle) Landau Hamiltonian; $B=(B_{\mu\nu})$ is
the matrix defining an external magnetic field. (Our conventions for
$B$ and $\theta$ differ from the usual ones by factors of 2, in order
to simplify some formulas. In particular, the magnetic field is $2B$.)
The interaction is
\eqa
\label{Hint} 
\cH_{\rm int} = g_0 \int d^{2n} \vx \, (\Psi^\dag\star\Psi\star \Psi^\dag\star\Psi)(\vx)
\eqaend
where $\star$ is the Groenewold-Moyal star product as usual (see e.g.\
Eq.\ (9) in \cite{LS}). The real coupling constant $g_0$ is arbitrary.

Formally, I define the quantum field theory by postulating standard
anticommutator relations for the fermion fields:
\eq 
\label{CAR0}
\Psi(\vx)\Psi^\dag(\vy)+ \Psi^\dag(\vy)\Psi(\vx) =
\delta^{2n}(\vx-\vy)
\eqend
etc. I will give a precise meaning to this model by introducing a
particular cutoff $\Lambda$ taking care of all potential
divergences. The limit $\Lambda\to\infty$ turns out to be rather trivial, at
least if one is only interested in the eigenfunctions and eigenvalues
of the Hamiltonian. In fact, we will be able to solve a more general
class of models with
\eqa 
\label{H0}
\cH_0 = \int d^{2n} \vx\, \Psi^\dag(\vx) (a H_{-B} + bH_{B} -\mu ) 
\Psi(\vx), \quad \eqaend
where we allow for an arbitrary linear combination of the Landau terms
with $B$ and $-B$ ($a,b>0$ are arbitrary constants), and I also find
it convenient to inserted a chemical potential $\mu$ (= arbitrary real
constant).  This generalization is interesting since, as is
well-known, the Landau Hamiltonian $H_B$ is highly degenerate, but in
this extended family of models we can also study the case where this
degeneracy is lifted. It is worth noting that, in the context of a
quantum Hall system, the second term has a natural physical
interpretation as a confining electrical potential.  As we will see,
there is only a single divergence in the model which can be removed by
normal ordering or, equivalently, an additive renormalization of
$\mu$.

We now rewrite our model by expanding the fermion fields in a
convenient basis of one particle wave functions. The resulting
Hamiltonian has a structure such that we can construct a complete set
of exact eigenstates and compute explicitly its partition function.
To simplify the presentation we first concentrate on the special case
$2n=2$ and then give the generalization to arbitrary dimensions $2n$.

\bigskip

\noindent {\em Remarks:} To prepare for the results in the next
Section it might be helpful to point out a complimentary
interpretation of the `noncommutative phase-space' model above: As
known since a long time, the star product corresponds to a
representation of Hilbert space operators by functions
\cite{BFFLS}. Thus the non-commutative field theory amounts to
replacing the one-particle states of the theory by Hilbert space
operators: plane waves are replaced by operator obeying
\eq
e^{i\vk \cdot \hat \vx} e^{i\vk'\cdot \hat \vx} = e^{i (\vk+\vk')\cdot \hat \vx} e^{ i \vk\cdot \theta \vk'} 
\eqend
where $\vk\cdot \theta \vk' = k_\mu \theta^{\mu\nu}k'_\nu$. The latter
relation is equivalent to interpreting the components of $\hat \vx$ as
operators obeying the commutator relations $[\hat x^\mu, \hat x^\nu] =
-2i\theta^{\mu\nu}$ (the Fourier variables $k_\mu,k'_\nu$ are real
numbers). Thus, instead of fields $\Psi(\vx)$ multiplied with the star
product, we could also use fields
\eq
\Psi(\hat \vx) = \int_{\R^{2n}}\frac{d^{2n} k}{(2\pi)^n} \tilde\Psi(k) \,
e^{i\vk\cdot \hat \vx} \eqend
with `standard' products (and the Fourier transform $\tilde\Psi$ an
ordinary function). Indeed, for $a=1$ and $b=0$ one could also write
the Hamiltonian above as
\eq 
\label{heuristic}
\cH = {\rm Trace}_{ L^2(\R^{2n}) } \left( \Psi^\dag(\hat \vx) 
\hat \vp^2\Psi(\hat \vx) + g [\Psi^\dag(\hat \vx)\Psi (\hat \vx)]^2 \right) 
\eqend
with 
\eq 
\label{CCR}
\hat x^\mu = x^\mu - i\theta^{\mu\nu}\partial_\nu,\quad \hat p_\mu
= -i\partial_\mu - B_{\mu\nu} x^\nu \eqend
the `Schr\"odinger representation' of the non-commutative field
theory. Eq.\ \Ref{CCR} implies \eq [\hat p_\mu,\hat x^\nu] =
-i(\delta_\mu^\nu + B_{\mu\lambda}\theta^{\lambda\nu}) , \eqend
i.e.\ if $B\theta=-I$ the `non-commutative positions' $\hat \vx$
commute with the `non-commutative Laplacian' $\hat \vp^2$! This
suggests that the model should be special at $B\theta=-I$, and in
particular it should have a huge gauge-like symmetry.

We also recall that, in a standard field theory, one can expand the
fields $\Psi$ in a basis $|\ell\rangle$ (Dirac-notation) of
eigenfunctions parametrized by positive integers $\ell$, and the
expansion coefficients are infinite vectors
$(A_\ell)_{\ell=1}^\infty$. The corresponding basis
$|\ell\rangle\langle m|$ for the Hilbert space operators is labeled by
two integers $\ell,m$, and we thus should expect that fields can be
represented by infinite matrices $(A_{\ell m})_{\ell,m=1}^\infty$.
This suggests a close relation of noncommutative field theories and
matrix models (various such relations have been discussed before
\cite{KS,DN,S}, but I believe the one used in the next Section
is particularly simple).

These remarks will be made precise in the next Section.

\bigskip 

\noindent {\bf 3. Matrix form of the $2+1$ dimensional model.} We
assume $2n=2$. My discussion will be based on two mathematical
facts. Both facts are known since many years in the context of phase
space quantization (a recent discussion including references to early
work can be found in \cite{CUZ}). More recently they have been used in
the context of noncommutative solitons \cite{GMS} (for other
references see \cite{KS,DN,S}). However, I would like to stress their
central importance for noncommutative field theory somewhat more than
usually done in this context (not only for the models discussed here
but in general). I have formulated them such that they are true, as
they stand, also in $2n$ dimensions (this is shown further below). For
the convenience of the reader, (elementary) proofs of these facts are 
given in Appendix~A.

\bigskip

\noindent {\bf Fact~1:} {\em There is a complete, orthonormal basis of
one-particle wave functions $\phi_{\ell m}(\vx)$, labeled by
positive integers $\ell$ and $m$, and which have the following
star product relations,
\eq
\label{fact1}
\phi_{\ell m}\star\phi_{\ell' m'} = \cc \delta_{m,\ell'}\phi_{\ell m'}
\eqend
with some positive constant $\cc$.\footnote{This is true in any
dimension $2n$. For completeness I quote $\cc^{-2} = (4\pi)^n \sqrt{
\det(\theta) }.$} Moreover, $\phi_{\ell m}^\dag =\phi_{m \ell}^\nd$.}

\bigskip

\noindent This suggests to expand the fermion fields in this basis,
\eq
\Psi(\vx) = \sum_{\ell,m} A^\nd_{\ell m}\phi_{\ell m}(\vx),\quad 
\Psi^\dag(\vx) = \sum_{\ell,m} A^\dag_{\ell m}\phi^\dag_{\ell m}(\vx),\quad  
\eqend
where the fermion operators $A^{(\dag)}_{\ell m}$ obey the usual anticommutator relations,
\eqa
\label{CAR}
A^\nd_{\ell m}A^\dag_{\ell' m'}+A^\dag_{\ell' m'}A^\nd_{\ell m}&=&\delta_{\ell,\ell'}\delta_{m,m'} \nonu
A^\dag_{\ell m}A^\dag_{\ell' m'}+A^\dag_{\ell' m'}A^\dag_{\ell m} &= & 0 .
\eqaend
A simple computation (using Fact~1 and $\int d^{2} \vx \,
\phi^\dag_{\ell m} \star \phi_{\ell' m'} (\vx) = \delta_{\ell,\ell'}
\delta_{m,m'}$) yields,
\eq 
\label{Hint1} 
\cH_{\rm int} = g \sum A^\dag_{m_1\ell_1}A^{}_{m_1\ell_2}
A^\dag_{m_2 \ell_2}A^\nd_{m_2 \ell_1} \equiv g \, \Tr (A^\dag A A^\dag A ) 
\eqend
where 
$g= g_0 \cc^2$.  
Here we interpreted the $A^\nd_{\ell m}$ as components of
a (infinite) matrix $A$ with adjoint $A^\dag$ defined as
$(A^\dag)_{\ell m} \equiv A^\dag_{m \ell}$, and $\Tr$ is the
usual matrix trace (sum of the diagonal). Thus
this basis $\phi_{\ell m}$ allows us to write the interaction in a
matrix form. We now observe that this basis has
a remarkable physical interpretation.

\bigskip 

\noindent {\bf Fact~2:} {\em The functions $\phi_{\ell m}$ are
common eigenfunctions of the Landau Hamiltonians $H_B$ and $H_{-B}$ in
\Ref{HL} for $B= \theta^{-1}$. The corresponding eigenvalues
$E_{\ell}$ and $E_m$ only depend on $\ell$ and $m$, respectively}. 

\bigskip

\noindent The latter property is the well-known degeneracy of the
Landau Hamiltonian. We also recall that the eigenvalues are identical
to those of a harmonic oscillator,
\eq
\label{2n2}
E_\ell=4 |B|(\ell-\frac12), 
\eqend
(recall that we label states by $\ell=1,2,\ldots$) and similarly for
$E_m$. Thus, if we choose the magnetic field as $B= \theta^{-1}$, the
free part of the Hamiltonian in \Ref{H0} has the following simple
form,
\eq
\label{H01}
\cH_0 = \sum_{m,\ell} ( E_{m} + \tilde E_\ell) 
A^\dag_{\ell m}A^\nd_{\ell m} 
\eqend
where we set $a=1$ and defined $\tilde E_\ell = b E_\ell -\mu$. The
latter is only to simplify notation: in our solution below the
explicit form of $E_m$ and $\tilde E_\ell$ is not needed.

\bigskip

\noindent {\em Remarks:} Fact~1 shows that the functions $\cc
\phi_{\ell m}$ provide a representation of rank-one Hilbert space
operators $|\ell\rangle\langle m|$ (Dirac bra-ket notation).  The
existence of such functions should not be surprising (see the remarks
at the end of the last Section). What I find remarkable, however, is
that these functions are old friends to anybody familiar with the
theory of the fractional quantum Hall effect (Fact~2) \cite{QH}.
\bigskip 

\noindent{\bf 4. Regularization and integrability.}  We consider the
eigenstates and eigenvalues of the field theory Hamiltonian
$\cH=\cH_0+\cH_{\rm int}$ defined in Eqs.\ \Ref{H01} and \Ref{Hint1},
with fermion operators obeying the usual anticommutator relations
\Ref{CAR}.  We use the regularization defined by restricting the
quantum numbers $\ell,m=1,2,\ldots \Lambda$ with
$\Lambda<\infty$. It is interesting to note that $\Lambda$ provides a
UV (= short-distance) and IR (= long-distance) cutoff at the same time
(the interested reader can find a more detailed discussion on this in
\cite{LS}).  The regularized model is defined on the fermion Fock
space $\cF_\Lambda$ generated by the fermion creation operators
$A_{\ell m}^{\dag}$, $1\leq \ell,m<\Lambda$, from a normalized vacuum
$\Omega$ defined by
\eq
A^\nd_{\ell m}\Omega=0 \quad \mbox{ for all $\ell,m$}, 
\eqend
and such that $\dag$ is the Hilbert space adjoint.  Since
$\cF_\Lambda$ is finite dimensional\footnote{$\cF_\Lambda\cong
\C^{\cN}$ with $\cN=2^{\Lambda^2}$.} all potential divergences are
taken care of by this regularization. 

It is convenient to introduce normal ordering $:\cdots :$ in the
interaction as usual.\footnote{
$
: A^\dag_{m_1\ell_1} A^\nd_{m_1\ell_2} A^\dag_{m_2\ell_2}
A^\nd_{m_2\ell_1} : \; = \, A^\dag_{m_2\ell_2} A^\dag_{m_1\ell_1}
A^\nd_{m_1\ell_2} A^\nd_{m_2\ell_1} .
$
} 
We observe that
\eq : \cH_{\rm int} : \; = \cH_{\rm int} - g\Lambda
\sum_{m,\ell} A^\dag_{m \ell} A^\nd_{m \ell}, \eqend 
i.e., normal ordering amounts to a shift of the chemical potential,
$\mu\to \mu - g\Lambda $. This shift diverges as
$\Lambda\to\infty$ and corresponds to a renormalization of the
chemical potential. After this renormalization
all eigenvalues and eigenstates have a well-defined limit
$\Lambda\to\infty$. To see that we recall the following  
natural basis in the fermion Fock space $\cF_\Lambda$,
\eq 
\label{vN} 
|N\rangle
\; = \; A^\dag_{\ell_1m_1} A^\dag_{\ell_2m_2}\cdots
A^\nd_{\ell_N m_N}\Omega \eqend
distinguished by the fermion number $N=0,1,2, \ldots $ and labeled by
$N$ distinct pairs $(\ell_j,m_j)$. As we make explicit below, the
Hamiltonian $\cH=\cH_0 + :\cH_{\rm int}:$ acting on all $|N\rangle$
is always well-defined even in the limit $\Lambda\to\infty$. Thus
{\em normal ordering is enough to remove all divergences in all the
eigenfunctions and eigenvalues of the model}.

In the following I write $\cH_{\rm int}$ short for $:\cH_{\rm int}:$.

\bigskip

\noindent {\em Remark:} The computation of other quantities of
physical interest may still require some regularization, of course.
For example, interpreting the model as a quantum Hall system, one
would be interested to study the model at fixed fermion density, i.e\
compute the partition function of the model at finite cut-off
$\Lambda$, and then take the limit $\Lambda\to\infty$ at fixed
expectation value of $N/\Lambda$.

\bigskip 

Before constructing the solution we give an indirect argument showing 
integrability of the model. For that we define the operators 
\eqa
\rho^\nd_{\ell m}= \sum_k  A^\dag_{k \ell} A^\nd_{k m}, \quad 
\tilde \rho^\nd_{\ell m}= \sum_k  A^\dag_{\ell k} A^\nd_{m k} 
\eqaend
and observe that they provide two commuting representations of the Lie
algebra $\gl_\infty$: using the fermion anticommutator relations one can 
show that the $\rho$'s obey the commutator relations
\eq [\rho_{\ell m}, \rho_{\ell' m'}] = \delta_{m,\ell'}\rho_{\ell m'}
- \delta_{m',\ell}\rho_{\ell' m} \: \eqend 
and similarly for the $\tilde\rho$'s, and $[\rho_{\ell m}, \tilde
\rho_{\ell' m'}]=0$ (the latter fact is not obvious 
but follows from a straightforward computation). Moreover,
\eq
\rho_{\ell m}^\dag = \rho_{m\ell}^\nd 
\eqend
and similarly for the $\tilde\rho$'s. We thus see that {\em the
operators $\rho$ and $\tilde\rho$ represent the Lie algebra
$\gl_\infty\oplus \gl_\infty$}. We also note
\eq
\label{hw1}
\rho_{\ell m}\Omega=\tilde \rho_{\ell m}\Omega=0 \quad \mbox{ for all $\ell,m$} . 
\eqend

\bigskip

{\em Remark:} For the readers appreciating fine points in analysis I
should stress that when I write $\gl_\infty$ I mean the inductive
limit $\Lambda\to\infty$ of $\gl_\Lambda$ with $\Lambda$ the matrix
cutoff described above: since we only consider the action of operators
on finite particle vectors $|N\rangle$, this limit is trivial.  Put
differently: it is true that some of our Fock space operators become
unbounded for $\Lambda\to\infty$, but we always consider them on a
particularly nice common dense invariant dense domain. (One could
without difficulty replace everywhere in our discussion `$\infty$' by
`$\Lambda$').

\bigskip

I now discuss the dynamical symmetry of the model. Note that the free
part of the model is a linear superposition of Cartan elements of
these representations, and the interaction is proportional to a
Casimir operator. More specifically, \Ref{H01} is equivalent to
\eq
\cH_0 = \sum_m ( E_m\rho_{mm} + \tilde E_m \tilde\rho_{mm}) , 
\eqend
and the interaction can be written in the following two equivalent
forms,
\eq  
\cH_{\rm int} \;   = g \sum_{\ell,m} : \rho_{\ell m}
\rho_{m\ell}: \, =\,  - g \sum_{\ell,m} : \tilde  \rho_{\ell m} \tilde
\rho_{m\ell}:   
\eqend
(the first equality here is obvious from \Ref{Hint1}, and the second
is obtained by interchanging the two $A$'s in \Ref{Hint1} using that
all fermion operators anticommute under the normal ordering
symbol). This shows that $\cH$ is a sum of commuting operators.
Moreover, even though the $\rho$'s and $\tilde\rho$'s commute with the
interaction $\cH_{\rm int}$, they do not commute with $\cH_0$:
\eqa
\label{rtr}
{[}\cH_0, \tilde \rho_{\ell \ell'}{]}&=&(\tilde E_\ell-\tilde E_{\ell'})
\tilde \rho_{\ell \ell'} \nonu
{[}\cH_0,\rho_{m m'}{]}&=&(E_m-E_{m'})\rho_{m m'} . 
\eqaend
This shows that the Lie algebra $\gl_\infty\oplus\gl_\infty$ is a
dynamical symmetry for our model. This rich symmetry structure
suggests that group theory should provide powerful tools to elegantly
solve this model. We also note the following commutator relations
\eqa
\label{rA} 
{[}\rho^\nd_{\ell'\ell''},A^\dag_{\ell m }{]}&=& \delta_{\ell'',\ell} A^\dag_{\ell' m } \nonu
{[}\rho^\nd_{m'm''},A^\dag_{\ell m }{]}&=& \delta_{m'',m} A^\dag_{\ell m'}  
\eqaend
which will be useful for us below.

We now compute the action of the Hamiltonian on the vectors
$|N\rangle$ defined in \Ref{vN}.  All $|N\rangle$ obviously are
eigenstates of the free Hamiltonian $\cH_0$ in \Ref{H01} with
eigenvalue
\eq 
\label{cE0} 
\cE_0 = \sum_{j=1}^N( \tilde E _{\ell_j} +  E_{m_j}) .  \eqend
Moreover, by a straightforward computation using the fermion
anticommutator relation we obtain the following equation
\eqa
\label{crux}
\cH_{\rm int} | N \rangle = 2 g \sum_{1 \leq j<k\leq N} T_{(jk)} | N \rangle 
\eqaend
with $T_{(jk)}$ a transposition operator defined as follows,
\eq
\label{Tt}
T_{(jk)} A^\dag_{\ell_1m_1} \cdots A^\dag_{\ell_N m_N}\Omega =
A^\dag_{\ell_1m_1} \cdots A^\dag_{\ell_jm_k}\cdots A^\dag_{\ell_km_j}
\cdots A^\dag_{\ell_Nm_N}\Omega \quad (j<k)  \eqend
(i.e.\ $m_j$ and $m_k$ are interchanged). Eqs.\ \Ref{crux}--\Ref{Tt}
will be the key to our solution.  Note that the operators $T_{(jk)}$
generate a (highly reducible) representation $T$ of the permutation
group $S_N$ which acts on the states $|N\rangle$ by permuting the
quantum numbers $m_j$. Further below we will show how to use the
representation theory of $S_N$ to construct eigenstates and
eigenvalues of the model. However, we first turn to a different
approach exploiting the dynamical symmetry.

\bigskip

\noindent {\em Remark:} A physical interpretation of the relations
\Ref{rtr}--\Ref{rA} is as follows. The one particle energies of our
model are sums of two parts, $E_{\ell m}=\tilde E_\ell +E_m$.  Since
${[}\cH_0,A^\dag_{\ell m}{]}=(\tilde E_\ell +E_m) A^\dag_{\ell m}$,
applying the fermion operator $A^\dag_{\ell m}$ adds the corresponding
one-particle energy to the state, as usual. The peculiar feature of
our model is that we have operators $\tilde\rho_{\ell \ell'}$ and
$\rho_{m m'}$ allowing to change only parts of the one-particle
energy. 

\bigskip

\noindent{\bf 5. Solution I. Pedestrian approach.}  We observe that we
can generate new eigenstates of $\cH$ from known ones by applying
operators $\tilde \rho_{\ell \ell'}$ and $\rho_{m m'}$: according to
our discussion above this gives new eigenstates where the eigenvalues
are changed by amounts $E_m-E_{m'}$ and $\tilde E_\ell-\tilde
E_{\ell'}$, respectively (this follows from \Ref{rtr}). We thus
can obtain many eigenstates from a sufficiently large number of
special ones.

We first construct special states of the form \Ref{vN} which are
eigenstates of {\em all} transpositions $T_{(jk)}$ and thus are
trivially also eigenstates of $\cH_{\rm int}$. Obvious such states are
those where all $m_j$ are the same, e.g.
$$
A^\dag_{1,1} A^\dag_{2,1}\cdots A^\dag_{N,1}\Omega 
$$
(note that due to the Pauli principle, i.e.\ $(A_{\ell
m}^\dag)^2=0$, such a state is non-zero only if all $\ell_j$ are
different): since all $m_j=m_k$, applying $T_{(jk)}$ does not change
anything, i.e.\ this is an eigenstate of all $T_{(jk)}$ with
eigenvalue equal to $1$. Thus this state is eigenstate of $\cH_{\rm
int}$ with eigenvalue $2g\vv$ where
$$
\vv= \left( \mbox{number of $T_{(jk)}$ with $1\leq j<k\leq N$} \right)  = 
\frac12 N(N-1).
$$
A more general such state is 
$$
|{[}\lambda_1,\lambda_2{]}\rangle = A^\dag_{1,1} A^\dag_{2,1}\cdots A^\dag_{\lambda_1,1}
A^\dag_{1,2} A^\dag_{2,2}\cdots A^\dag_{\lambda_2,2} \Omega,
$$
with $\lambda_1+\lambda_2=N$, i.e.\ we have two groups of fermion creation
operators where the $m_j$ are equal within each group but different in
the other group. This is obviously an eigenstate of all $T_{(jk)}$
with $j$ and $k$ both in the same groups (i.e.\ $1\leq j<k\leq \lambda_1$
and $\lambda_1+1\leq j<k\leq \lambda_1+\lambda_2$), and the eigenvalue of all
these is $+1$. Moreover, it is also in an eigenstate of
$T_{1,\lambda_1+1}$ but with eigenvalue equal to $-1$: $T_{1,\lambda_1+1}$
interchanges $A^\dag_{1,1}$ and $A^\dag_{1,2}$, but due to the fermion
anticommutator relations this is the same as multiplying with $-1$.
Similarly, all $T_{j,\lambda_1+j}$ have eigenvalues $-1$. However,
applying $T_{1,\lambda_1+2}$ changes $A^\dag_{1,1}$ to $A^\dag_{1,2}$ and
$A^\dag_{2,2}$ to $A^\dag_{2,1}$. Now at least one of the
operators $A^\dag_{1,2}$ or $A^\dag_{2,1}$ appears twice, and the
resulting state thus is zero (Pauli principle).  The same is true
for all other $T_{(jk)}$ with $j$ in the first and $k$ in the second
group and $\ell_k\neq \ell_j$: all these annihilate the state
$|{[}\lambda_1,\lambda_2{]}\rangle$ due to the Pauli principle.  We thus
conclude that $|{[}\lambda_1,\lambda_2{]}\rangle$ is an eigenstate of
$\cH_{\rm int}$ with eigenvalue $2g\vv$ where
$$
\vv=   \frac12\lambda_1(\lambda_1 -1) +  \frac12 \lambda_2(\lambda_2-1) -\min(\lambda_1,\lambda_2)  
$$
(we counted the number of $T_{(jk)}$ belonging to the same group and
subtracted the number of $T_{(jk)}$ belonging to different groups but
with $\ell_k = \ell_j$). It now is obvious how to generalize this to
states with an arbitrary number $L=1,2,\ldots N$ different groups: The
crucial point in the previous example was not only that the $m_j$ in
each group coincide, but also that a maximal number of the $\ell_j$ in
the different groups are the same. We thus define
\eq
\label{bnu}
|{[}\lambda{]}\rangle = A^\dag_{1,1} A^\dag_{2,1}\cdots
A^\dag_{\lambda_1,1} A^\dag_{1,2} A^\dag_{2,2}\cdots
A^\dag_{\lambda_2,2}\cdots A^\dag_{1,L} A^\dag_{2,L}\cdots
A^\dag_{\lambda_L,L} \Omega
\eqend
with a set ${[}\lambda{]}={[}\lambda_1,\lambda_2,\ldots,\lambda_L{]}$ of positive
integers $\lambda_i$ such that $\lambda_1+\lambda_2+\ldots + \lambda_L = N$.
Similarly as above we check that this is an eigenstate of all $T_{jk}$
with eigenvalue $+1$ if both $j,k$ belong to the same group (i.e.,
$m_j=m_k$), $-1$ if $j,k$ belong to different groups but
$\ell_j=\ell_k$, and $0$ in all other cases. Thus $|{[}\lambda{]}\rangle$ is
an eigenstate of $\cH_{\rm int}$ with eigenvalue $2g\vv_{{[}\lambda{]}}$ where  
$$
\vv_{{[}\lambda{]}} = 
\sum_{i=1}^L \frac12\lambda_i(\lambda_1-1) - \sum_{1\leq i<j\leq L}\min(\lambda_i ,\lambda_j) 
$$
We thus have obtained various particular eigenstates of $\cH$ with
different eigenvalues of the interaction $\cH_{\rm int}$. To avoid
confusion we note that each of these states represents a whole class
of states, e.g.\ using \Ref{rA} we can write any eigenstate of $\cH$
where all $m_j$ are the same as follows,
$$
A^\dag_{\ell_1 m} A^\dag_{\ell_2 m}\cdots A^\dag_{\ell_N m}\Omega
= \tilde\rho_{\ell_1,1}\tilde\rho_{\ell_2,2}\ldots \tilde\rho_{\ell_N,N}
(\rho_{m,1})^N |{[}N{]}\rangle , 
$$
where $|{[}N{]}\rangle= A^\dag_{1,1} A^\dag_{2,1}\cdots
A^\dag_{N,1}\Omega$ is the particular state which for now is the
only one of this kind taken into account.
In fact, at this stage we can further restrict ourselves to states
$|{[}\lambda{]}\rangle$ which are in one-to-one correspondence to partitions
${[}\lambda{]}={[}\lambda_1,\lambda_2,\ldots,\lambda_N{]}$ of $N$,
\eq \lambda_1\geq \lambda_2\geq \ldots \geq \lambda_L > 0,\quad
\lambda_1+\lambda_2+\cdots +\lambda_L=N . 
\eqend
This allows us to write a somewhat simpler formula for the
possible eigenvalues of $\cH_{\rm int}$,
\eq
\label{vv} 
\vv_{{[}\lambda{]}} = 
\sum_{i=1}^L \frac12\lambda_i(\lambda_i+1) - \sum_{i=1}^L i \lambda_i  . 
\eqend

All these states $|{[}\lambda{]}\rangle$ are obviously also eigenstates of
$\cH_0$ with eigenvalue $\bcE_0\equiv \cE_0$ as in \Ref{cE0} but with
$\ell_j=\bar{\ell}_j$ and $m_j=\bar{m}_j$ where
\eqa 
\label{lm00}
(\bar{\ell}_1,\bar{\ell}_2,\ldots,\bar{\ell}_N) =
(1,2,\ldots,\lambda_1, 1,2,\ldots,\lambda_2,\ldots, 1,2,\ldots,\lambda_L ) \nonu
( \bar{m}_1,\bar{m}_2,\ldots,\bar{m}_N)  =
(\underbrace{1,1,\ldots,1}_{\lambda_1}, \underbrace{2,2,\ldots,2}_{\lambda_2}
,\ldots, \underbrace{L,L,\ldots,L}_{\lambda_L} ) .  \eqaend
Using this notation we can also write
\eq
\label{bnu1}
|{[}\lambda{]}\rangle = 
A^\dag_{\bar{\ell}_1 \bar{m}_1} A^\dag_{\bar{\ell}_2 \bar{m}_2}
\cdots A^\dag_{\bar{\ell}_N \bar{m}_N}\Omega 
\eqend
Thus $|{[}\lambda{]}\rangle$ is an eigenstate of $\cH$ with eigenvalue
$\bcE = \bcE_0+2g \vv_{{[}\lambda{]}}$. As discussed above, we can now
generate other eigenstates with different eigenvalues by applying
operators $\tilde \rho_{\ell \ell'}$ and $\rho_{m m'}$ to known
eigenstates. In particular, the state
\eqa
\label{Phi}
\Phi= 
\tilde \rho_{\ell_{1},\bar{\ell}_1} \tilde \rho_{\ell_{2},\bar{\ell}_2}
\cdots\tilde \rho_{\ell_{N},\bar{\ell}_N}
\rho_{m_{1},\bar{m}_1} \rho_{m_{2},\bar{m}_2}\cdots 
\rho_{m_{N},\bar{m}_N} 
|{[}\lambda{]}\rangle 
\eqaend
is an eigenstate of $\cH$ but with eigenvalue
\eq 
\label{cE}
\cE= \cE_0 + 2g \vv_{{[}\lambda{]}}, \quad \mbox{$\cE_0$ in \Ref{cE0} and
$\vv_{{[}\lambda{]}}$ in \Ref{vv} } \eqend
(to see this, apply $\cH_0$ to $\Phi$ in \Ref{Phi}, move it to the
right using repeatedly the relations in \Ref{rtr}, and note that the
energy differences thus generated add up to $\cE_0-\bcE_0$.) It is
worth noting that the second of the quantum numbers for the operators
$\rho$ and $\tilde\rho$ in \Ref{Phi} were forced on us by Eqs.\
\Ref{rA} and \Ref{hw1}: any other choice would have given zero (this
will become obvious further below). I will argue below that we thus
have obtained all eigenstates and eigenvalues of our model. In fact,
we obtained too many: many of the states in \Ref{Phi} are actually
actually zero and thus not all of the eigenvalues in \Ref{cE} exist.
For example, if all $m_j=m$ are identical, $\Phi$ in \Ref{Phi} is
nonzero only for ${[}\lambda{]}={[}N{]}$, i.e.\ in this case only the
eigenvalue $\cE$ \Ref{cE} exists where $\vv =N(N-1)/2$. To
characterize all non-zero states is a non-trivial combinatorial
problem which we will discuss in more detail further below. In the
rest of this Section we derive a more explicit formula of the
eigenstates $\Phi$ \Ref{Phi}.

It is useful to first consider the special case ${[}\lambda{]}={[}N{]}$,
i.e. 
$$
\Phi = \tilde \rho_{\ell_{1},1} \tilde \rho_{\ell_{2},2} \cdots\tilde
\rho_{\ell_{N},N} \rho_{m_{1},1} \rho_{m_{2},1}\cdots \rho_{m_{N},1}
A^\dag_{1,1} A^\dag_{2,1}\cdots A^\dag_{N,1}\Omega.
$$
We commute $\rho_{m_{N},1}$ to the right of all $A^\dag$'s using
repeatedly \Ref{rA}, until it hits $\Omega$ an disappears (gives zero
according to \Ref{hw1}).  This produces a sum of $N$ terms, and in
each of them one of the $\bar{m}_j=1$ is changed to $m_N$. Similarly
we then remove $\rho_{m_{N-1},1}$, which turns in every term one of
the remaining $\bar{m}_j=1$ to $m_{N-1}$, etc., until we remove
$\rho_{m_{1},1}$ and the remaining $\bar{m}_j=1$ in each term is turned
to $m_1$. Thus we used \Ref{rA} and \Ref{hw1} to remove all
$\rho$'s, which turns all $\bar{m}_j=1$ to $m_j$, and we obtain $N!$
different terms where we have the $m_j$ in all possible different
orders. The result after these manipulations can be written as
$$
\Phi = \tilde \rho^\nd_{\ell_{1},1} \tilde
\rho^\nd_{\ell_{2},2} \cdots\tilde \rho^\nd_{\ell_{N},N} 
\ccS_{1,2,\ldots , N}A^\dag_{1,m_1} A^\dag_{2,m_2}\cdots
A^\dag_{N,m_N}\Omega  
$$
where $\ccS_{1,2,\ldots ,N}$ means symmetrization over all
indices $m_j$.  In a similar manner we can remove all
$\rho_{\ell_{j},j}$, and this now turns the indices
$\bar{\ell}_j=j$ to $\ell_{j}$. Since all $\bar{\ell}_j$ are
different and equal to only one of the corresponding indices of the
$A^\dag$, the number of terms does not increase and we obtain
\nonueqa
\Phi = 
\ccS_{1,2,\ldots , N}  A^\dag_{\ell_1 m_1} A^\dag_{\ell_2 m_2}\cdots
A^\dag_{\ell_N m_N}\Omega  = \\
\sum_{P\in S_N } 
 A^\dag_{\ell_1 m_{P(1)}} A^\dag_{\ell_2 m_{P(2)}}\cdots
A^\dag_{\ell_N m_{P(N)}} \Omega 
\nonueqaend
which is the final formula for the eigenstate in this special case. 

It is easy to generalize this to the general case by inserting
\Ref{lm00} and \Ref{bnu1} into Eq.\ \Ref{Phi}. Again we can remove all
the $\rho$'s and $\tilde\rho$'s using \Ref{rA} and \Ref{hw1}. Removing
the $\rho$'s turns all $\bar{m}_j$'s to $m_j$'s, but we obtain
$\lambda_1!\times \lambda_2!\times \cdots \times \lambda_L!$ terms
corresponding to how often the numbers $1,2,\ldots,L$ appear in $\{
\bar{m}_j\}_{j=1}^N$ in \Ref{lm00}: $1$ appears $\lambda_1$ times and
thus one has symmetrization in the first $\lambda_1$ indices $m_j$,
$2$ appears $\lambda_2$ times and thus one has symmetrization in the
next $\lambda_2$ indices $m_j$, etc. Similarly, removing the $\tilde
\rho$'s turns the $\bar{\ell}_j$'s to $\ell_j$'s, but we further
increase the number of terms by factors $\tilde\lambda_1!\times
\tilde\lambda_2!\times \cdots \times \tilde\lambda_{N}!$ where
$\tilde\lambda_i$ is the number of times $i$ appears in
$\{\bar{\ell}_j \}_{j=1}^N $.  The symmetrizations are now in the
indices $\ell_j$ at the positions $j$ where the $\bar{\ell}_j$
coincide. The numbers $\tilde\lambda_i$ define another partition
${[}\tilde
\lambda{]}={[}\tilde\lambda_1,\tilde\lambda_2,\ldots,\tilde\lambda_{\tilde
L}{]}$ of $N$, e.g.\ if ${[}\lambda{]}={[}5,3,2 {]}$ then ${[}\tilde
\lambda{]}={[}3,3,2,1,1{]}$ (readers familiar with Young tableaux will
recognize this as the conjugate partition \cite{Ro}). The resulting
formula for $\Phi$ \Ref{Phi} can be written conveniently in terms of
partial symmetrization operators,
\eqa \ccS_{j_1,j_2,\ldots,j_K} A^\dag_{\ell_1 m_1} A^\dag_{\ell_2 m_2}\cdots A^\dag_{\ell_N m_N}\Omega 
\, : = 
\nonu \sum_{P\in S_K} 
A^\dag_{\ell_1 m_1} \cdots A^\dag_{\ell_{j_1} m_{j_{P(1)}
}}\cdots A^\dag_{\ell_{j_2} m_{j_{P(2)} }}\cdots  
A^\dag_{\ell_{j_K} m_{j_{P(K)} }}\cdots A^\dag_{\ell_N m_N} \Omega 
\eqaend
(this sum of $K!$ terms
corresponds to symmetrization in the $K$ indices $m_j$ at the
positions indicated by the subscripts), and another similar
symmetrization but with respect to the indices $\ell_j$,
\eqa \tcS_{j_1,j_2,\ldots,j_K} A^\dag_{\ell_1 m_1}
A^\dag_{\ell_2 m_2}\cdots A^\dag_{\ell_N m_N}\Omega \, := \nonu
 \sum_{P\in S_K} A^\dag_{\ell_1 m_1} \cdots
A^\dag_{\ell_{j_{P(1)} m_{j_1} }}\cdots
A^\dag_{\ell_{j_{P(2)} m_{j_2} }}\cdots
A^\dag_{\ell_{j_{P(K)} m_{j_K} }}\cdots A^\dag_{\ell_N m_N}
\Omega .  \eqaend
The general formula is 
\eqa\label{Phi2}
\Phi = 
\ccS_{1,2,\ldots,\lambda_1} \ccS_{\lambda_1+1,\lambda_1+2, \ldots\lambda_1+\lambda_2}\ldots
\ccS_{\lambda_1+\ldots+\lambda_{N-1}+1,\ldots, N}
\times \nonu \times 
\tcS_{1,\lambda_1+1,\lambda_1+\lambda_2+1,\ldots,\lambda_1+\ldots + \lambda_{\tilde\lambda_1}+1}
\tcS_{2,\lambda_1+2,\ldots,\lambda_1+\ldots + \lambda_{\tilde\lambda_2}+2}\ldots
\tcS_{\lambda_1,\ldots,\lambda_1+\lambda_2+\lambda_{\tilde\lambda_{\tilde L}}+\lambda_1}|N\rangle 
\eqaend 
with $|N\rangle$ defined in \Ref{vN}. To digest this general formula
is is useful to consider few special cases, e.g.\
\eq
\label{Ex}
\Phi = \ccS_{1,2,3,4,5}
\ccS_{6,7,8} \ccS_{9,10}
\tcS_{1,6,9} \tcS_{2,7,10}\tcS_{3,8}|N\rangle \quad \mbox{ for
${[}\lambda{]}={[}5,3,2{]}$.}
\eqend
(we used $\tcS_{4}=\tcS_{5}=I$) etc. A simple method to determine the
subscripts in \Ref{Phi2} is to draw the Young tableaux corresponding
to the partition ${[}\lambda{]}$ but write the numbers $1,2,\ldots,N$ instead
of the usual boxes, in increasing order from left to right and from up
to down. For our example this gives
$$
\bma{ccccc} 1&2 & 3& 4& 5 \\ 6 & 7& 8& & \\ 9 & 10& & &
\ema \quad \mbox{for ${[}\lambda{]}={[}5,3,2{]}$.}
$$
For each column write one $\tcS$ and for each row one $\ccS$, and the
subscripts of the $\tcS$ in \Ref{Phi2} are given by the numbers in the
columns and the subscripts of the $\ccS$ by the numbers in the rows.
(The order in which the $\tcS$'s or the $\ccS$'s are written is
irrelevant, of course.)

To make contact with the results obtained in the next Section we note
that the interaction can be written also as sum of transpositions
$\tilde T_{(jk)}$ in another representation $\tilde T$ of the
permutation group $S_N$: Defining
\eq
\label{tTt}
\tilde T_{(jk)} A^\dag_{\ell_1m_1} \cdots A^\dag_{\ell_N m_N}\Omega =
A^\dag_{\ell_1m_1} \cdots A^\dag_{\ell_k m_j}\cdots A^\dag_{\ell_j m_k}
\cdots A^\dag_{\ell_Nm_N}\Omega \quad (j<k) 
\eqend
(i.e.\ $\ell_j$ and $\ell_k$ are interchanged) we observe that $\tilde
T_{(jk)}T_{(jk)}=-I$ (since this amounts to interchanging
$A^\dag_{\ell_j m_j}$ and $A^\dag_{\ell_k m_k}$ which is the same as
multiplication with $-1$). We thus can also write
\eqa
\label{cruxt}
\cH_{\rm int} | N \rangle = -2 g \sum_{1 \leq j<k\leq N} \tilde
T_{(jk)} | N \rangle . \eqaend
We thus have two representations $T$ and $\tilde T$ of $S_N$ in our
model which are, however, not independent but such that their product
equals the representation $P\to (-1)^{|P|}$ (= parity of
$P\in S_N$). This fact allows us to somewhat simplify Eq.\
\Ref{Phi2}: we can replace the symmetrizers with respect to the
indices $\ell_j$ by antisymmetrizers with respect to the indices
$m_j$, i.e., the symmetrizers $\tcS$ are equal to 
\eqa \label{tS2} 
\tcS_{j_1,j_2,\ldots,j_K} A^\dag_{\ell_1 m_1}
A^\dag_{\ell_2 m_2}\cdots A^\dag_{\ell_N m_N}\Omega \, : = \nonu
\frac1{K!} \sum_{P\in S_K} (-1)^{|P|} A^\dag_{\ell_1 m_1} \cdots
A^\dag_{\ell_{j_{1} } m_{ j_{j_P(1)} }}\cdots
A^\dag_{\ell_{j_{2} } m_{ j_{j_P(2)} }}\cdots
A^\dag_{\ell_{j_{K} } m_{ j_{j_P(K)} }}\cdots
A^\dag_{\ell_N,m_N}
\Omega .  
\eqaend
Inserting that, we write $\Phi= Y^{{[}\lambda{]}}|N\rangle$. We now
observe that $Y^{{[}\lambda{]}}$ is, up to normalization, the Young
operator \cite{Ro} (or Young symmetrizer \cite{W}) associated with
${[}\lambda{]}$: this result will be obtained in the next section
using group theory.

\bigskip

\noindent {\em Remark:} It is interesting to note that the model
remains solvable if we add to the Hamiltonian $\cH=\cH_0 + \cH_{\rm
int}$ Hartree-Fock type terms of the following form,
\eq 
\label{HF}
\cH_{\rm HF} = 
\sum_{\ell,m} :\left(  V_{\ell m}\rho_{\ell\ell} \rho_{mm} + 
\tilde V_{\ell m} \tilde\rho_{\ell\ell}\tilde\rho_{mm} +
W_{\ell m}\rho_{\ell\ell}\tilde\rho_{mm} \right) :  \eqend
where $\tilde V,V$ and $W$ are arbitrary model parameters: the
eigenstates $\Phi$ in Eq.\ \Ref{Phi} remain the same, and the
eigenvalues $\cE$ are changed by adding $\sum_{i\neq j}( V_{\ell_i
\ell_j} + \tilde V_{m_i m_j} + W_{\ell_i m_j})$. We will come back to
that in Section~9.

\bigskip

\noindent{\bf 6. Solution II. Group theory approach.} As mentioned, there is a 
representation $T$ of the permutation group $S_N$ on the $N$-particle
states of the model, 
\eq
\label{T}
T_P
A^\dag_{\ell_1 m_1} A^\dag_{\ell_2 m_2}\cdots
A^\dag_{\ell_N m_N}\Omega = 
A^\dag_{\ell_1 m_{P(1)}} A^\dag_{\ell_2 m_{P(2)}}\cdots
A^\dag_{\ell_N m_{P(N)}}\Omega 
\eqend
for all $P\in S_N$. 

As we now discuss, diagonalizing $\cH$ amounts to decomposing $T$ in
irreducible representations (irreps).  We first recall \Ref{crux}: the
interaction applied to any state $|N\rangle$ defined in \Ref{vN} is
proportional to the sum of all transpositions,
\eq
\label{CN}
C_N = \sum_{1 \leq j<k\leq N} T_{(jk)} 
\eqend
to this state. The crucial
fact allowing to solve the model is that the free Hamiltonian $\cH_0$
commutes with all permutations $T_P$ in \Ref{T} (since the
eigenvalues $\cE_0$ in \Ref{cE0} are invariant under all permutations
of the indices $m_j$). If we therefore make an ansatz
\eq
\label{Phi1}
\Phi = \sum_{P\in S_N} a_P T_P |N\rangle
\eqend
(with $|N\rangle$ short for any of the vectors in \Ref{vN}) 
and choose the real coefficients $a_P$ so that
\eq
\label{ev} 
C_N \sum_{P\in S_N} a_P T_P 
= \vv  \sum_{P\in S_N} a_P T_P 
\eqend 
for some real $\vv$, then $\Phi$ is an eigenvector of $\cH$ with
eigenvalue $\cE=\cE_0 +2 g \vv$. 

Eq.\ \Ref{ev} can be interpreted as an eigenvalue equation for $C_N$
and can be solved as such. There are two obvious solutions: $a_P=1$
with $\vv=N(N-1)/2$ and $a_P=(-1)^{|P|}$ with $\vv=-N(N-1)/2$.  They
correspond the the eigenstates
\eq
|N\rangle_\pm = \sum_{P\in S_N} (\pm 1)^{|P|}  
T_P |N\rangle
\eqend
with the corresponding energy eigenvalues
\eq \cE_\pm = \cE_0 \pm N(N-1) g .  \eqend 
For fixed $\cE_0$ these are the extreme eigenvalues. The total number
of solutions is $N!$ generically. To appreciate the problem in Eq.\
\Ref{ev} it is instructive to solve it by brute-force as discussed in
Appendix~B, but this approach is only possible for small $N$.

To see that Eq.\ \Ref{ev} is equivalent to a classical group theory
problem, note that $C_N = \sum_{i<j} T_{(ij)}$ is a class operator,
i.e.\ it commutes with all permutations $T_P$. Thus $C_N$ is equal to
a constant $\vv$ in each irreps. Eq.\ \Ref{ev} thus amounts to
decomposing the representation $T$ into irreps.  The irreps of $S_N$
are well-known. They are in one-to-one with the partitions
${[}\lambda{]}$ of $N$, and the value $\vv_{{[}\lambda{]}}$ of $C_N$
in this irreps is well-known and equal to what we found in Eq.\
\Ref{vv} before (see e.g.\ Eq.\ (4-3) in \cite{Chen}). If all $\ell_j$
and all $m_j$ are different, then the multiplicity of these eigenvalue
equals $(k_{{[}\lambda{]}})^2$ where
\eq
\label{klambda}
 k_{{[}\lambda{]}} = \frac{N!\prod_{1\leq i<j\leq
L}(h_i-h_j)}{h_1!h_2!\cdots h_L!}, \quad h_i=\lambda_i+L-i \eqend
is the dimension the irreps ${[}\lambda{]}$ (this follows from the
fact that $T$ in this case is equivalent to the regular representation
of $S_N$ --- see e.g.\ Theorem 3.25 and Eq.\ (4-4a) in \cite{Chen};
the interested reader can find some examples in Appendix~B). The
corresponding $\sum_P a_P T_P$ is equal to the Young operator
$Y^{{[}\lambda{]}}$ \cite{Ro,W}.  We thus have recovered the result in
the previous Section.  The group theory argument is somewhat more
powerful in that it also shows that no eigenfunction was
missed. Moreover, the Young operators are projections up to
normalizations: these normalizations are known and give the
normalizations of the eigenstates.

\bigskip

\noindent {\em Remark:} We thus found all eigenstates and eigenvalues
of the model. One remaining problem is that we actually found too
many: For a given $|N\rangle$, we obtained $N!$ eigenfunctions
$Y^{{[}\lambda{]}}|N\rangle$, but all of them are linearly independent
only if all $N$ $\ell_j$'s and all $N$ $m_j$'s are different. E.g.\ if
all $m_j=m$ are equal, then only one of these eigenvectors is non-zero
(the others will all vanish due to the fermion anticommutator
relations) and similarly for states where all $\ell_j$ are equal. If
we put identical $\ell_j$'s in groups then the $\ell$-degeneracies
can be characterized by the numbers $\nu_j$ of elements in the different
groups. This defines a partition ${[}\nu{]}$. For example, for
$(\ell_j) = (2,5,5,2,3,2,5,2)$ one gets ${[}\nu{]}= {[}4,3,1{]}$.
Similarly the degeneracies of the $m_j$'s can be characterized by
another partition ${[}\mu{]}$.  Thus the degeneracies of the eigenvalues
$\cE$ depend on three partitions.  To find all these
multiplicities seems like a rather non-trivial exercise in
combinatorics. If one knew these multiplicities $m$, one could compute
the partition functions $\cZ={\rm Trace}_{\cF}\exp(-\beta\cH)$ of the
model as follows,
\eq \cZ = \sum_{(\nu),(\mu),[\lambda]} e^{-\beta \sum_{\ell=0}^\infty
( \nu_\ell \tilde E_{\ell} + \mu_\ell E_\ell ) }
m_{{[}\nu{]}{[}\mu{]}}^{{[}\lambda{]}}e^{-\beta 2 g \vv_{{[}\lambda{]}} }
\eqend 
where $(\nu)=(\nu_1,\nu_2,\ldots)$ with $\nu_\ell\geq 0$ and $[\nu]$
is the corresponding partition with the $\nu_i$'s ordered (to have
this sum finite one might still need a finite cut-off $\Lambda$, of
course).

\bigskip

\noindent {\bf 7. Generalization to $2n$ dimensions.}  We now discuss
the generalization to $2n$ dimensions. There one has (generalized)
Landau eigenfunctions $\phi_{\vl,\vm}$ which are labeled by $2n$
positive integers, $\vl=(\ell_1,\ldots,\ell_n)$ and similarly for
$\vm$.  This is obvious in the coordinate system where the matrix $B$
has Jordan normal,
\eq
\label{B}
\left(B_{\mu\lambda}\right)=\pmatrix{0&B_1& & & \cr-B_1&0& & & \cr
& &\ddots& & \cr & & &0&B_n\cr & & &-B_n&0\cr} \ , \quad B_\mu >0 , 
\eqend
so that $H_{B}$ in \Ref{HL} is the sum of $n$ terms $H_{B_j}$
depending only on the coordinates $x_{j-1}$ and $x_{2j}$,
$j=1,\ldots,n$. Thus $\phi_{\vl,\vm}(\vx)$ is just the product of $n$
two dimensional Landau eigenfunctions
$\phi_{\ell_j,m_j}(x_{2j-1},x_{2j})$, and the corresponding
eigenvalues of $H_B$ are 
\eq
E_\vl=\sum_{j=1}^n 4 B_j(\ell_j - \frac12).
\eqend
Then all what we discussed for two dimensions goes through with
$\ell,m$ replaced by $\vl,\vm$ etc. We now observe that we can map the
vectors $\vl=(\ell_1,\ldots,\ell_n)$ in a one-to-one way to a single
positive integer $\ell$ (e.g.\ for $n=2$ one such map is $(1,1)\to
1$, $(1,2)\to 2$, $(2,1)\to 3$, $(1,3)\to 4$, $(2,2)\to 5$, \ldots
$(\ell_1,\ell-\ell_1)\to \ell(\ell-1)/2+\ell_1$).  Doing that, all
formulas obtained in two dimensions hold true as they stand also in
$2n$ dimensions with the only difference that the eigenvalues $E_\ell$
and $E_m$ are given by somewhat more complicated expressions. However,
{\em our construction of eigenstates and eigenvalues above does not
rely on the explicit form of these eigenvalues, and we thus have
obtained all eigenstates and corresponding eigenvalues for
arbitrary dimensions $2n$.}

\bigskip

\noindent {\bf 8. More general solvable models:} An obvious
generalization of our model would be to add a further interactions
corresponding to a higher Casimir of $\gl_\infty$, for example the
term $\cH^{(3)} = \, g^{(3)} :\Tr(\rho^3) : $, i.e.
\eq : \cH^{(3)}_{\rm int}:\,  = \, g^{(3)} \sum_{k,\ell,m} : \rho_{k\ell}
\rho_{\ell m} \rho_{mk} :  \eqend 
with the colons indicating normal ordering as usual (i.e.\ move all
$A^\dag$ to the right of all $A$'s).  Again we can write this
interaction also in terms of the $\tilde \rho$'s, $\cH^{(3)} = \,
g^{(3)} :\Tr(\tilde \rho^3)$.  This term comes the following 3-body
interaction,
\eq \cH^{(3)}_{\rm int} = g^{(3)}_0 \int d^{2n} \vx \,
(\Psi^\dag\star\Psi\star \Psi^\dag\star\Psi\star
\Psi^\dag\star\Psi)(\vx) \eqend
where regularization by normal ordering amounts to a renormalization of
the 2-body interaction constant $g$ and the chemical potential
$\mu$. Applying this interaction to a state $|N\rangle$ \Ref{vN}
yields
\eq
:\cH^{(3)}_{\rm int}: |N\rangle = g^{(3)} \sum T_{(j k \ell)}|N\rangle  
\eqend
where the sum on the r.h.s.\ is over all 3-cycles of the permutation
group $S_N$ \cite{Mu}. The eigenstates of the model which we
constructed can be chooses such that they are also eigenstates of this
interaction. The possible eigenvalues $\vv^{(3)}$ of this are known:
they are characterized by partitions ${[}\lambda{]}$ and are given by (see
e.g.\ Eq.\ (4-3) in \cite{Chen})
\eqa \vv^{(3)} = 
\frac13\left\{ 
2N -\frac32N^2 +\sum_{i= 1}^N \lambda_i
\left[ \lambda_i^2 -\left(  3i-\frac32  \right)\lambda_i +3i(i-1) 
\right]
\right\}
\eqaend 
One can can add further interactions $\cH^{(p)}_{\rm int}\propto \;
:\Tr(\rho^p):$, $p=4,5,\ldots $, and still have a solvable model. It
is interesting to note that
\eq :\Tr(\rho^p): \, = (-1)^{p-1} :\Tr(\tilde\rho^p): .  \eqend

\bigskip

\noindent {\bf 9. Outlook and open questions.} 

\begin{itemize} 

\item {\it More general models.} The model solved in this paper is
only the simplest in a class of similar models: one can add a flavor
(or spin) index to the fermion operators and thus increase the
symmetry from $\gl_\infty$ to $\gl_k\otimes \gl_\infty$. This allows
for additional types of interactions (spin-spin-like for $k=2$,
e.g.). Models of this kind can be obtained, e.g., by truncating the 2D
Hubbard model: simplifying the 2D Hubbard interaction by keeping only
particular terms \cite{EL}. In 2D one can keep, in addition to
Hartree- and Fock-terms (leading to mean field theory), also
particular `mixed' terms and still have a model which, as I believe,
is exactly solvable \cite{EL}.

\item {\it Mean field vs.\ correlations.} To put my results in
perspective, I recall a well-known class of exactly solvable models
which can be defined by the Hamiltonian $\cH=\cH_0+\cH_{\rm HF}$ with
the free part in \Ref{H01} and the Hartree-Fock interaction in
\Ref{HF} (this is only a special case \cite{EL}). For this model all
states $|N\rangle$ in \Ref{vN} are eigenstates: Hartree-Fock
interactions only change the energy eigenvalues but not the energy
eigenstates. These eigenstates are Slater determinants: no
correlations. For the models $\cH=\cH_0+\cH_{\rm int}+\cH_{\rm HF}$
the eigenstates are highly non-trivial linear combinations of Slater
determinants (generically: there are exceptions, of course): for fixed
$\ell_j$'s and $m_j$'s one has (generically) many different Slater
determinants where the $\ell_j$'s and $m_j$'s are distributed over the
fermions in all kinds of different ways, and only very particular
linear combinations of these are energy eigenstates.  For example, if
all $\ell_j$'s and all $m_j$'s are different, the $10$-particle
eigenstate in Eq.\ \Ref{Ex} is a particular sum of $5!3!2!3!3!2!>10^5$
terms! I believe that most of these states cannot be written as Slater
determinants (a proof of this would be very welcome). Thus, different
from Hartree-Fock models, our model should describe correlated
fermions.

\item {\it Phase transitions?} In this paper I only demonstrate how to
solve the models. To explore the physics of the solution is left to
future work. However, to give a glimpse in that direction, I now
discuss one special case of physical interest in which one can see by
simple means that the solution can describe interesting physics:
Assume $n=2$, $a=1$ and $b=0$:
\eq
\cH = \sum_{m\geq 1} |B| (m-1)\rho_{mm}  + 
g\sum_{\ell,m}:\rho_{\ell m}\rho_{m \ell}:  
\eqend
(I set the chemical potential $\mu$ to some convenient value).  As
discussed, this is a toy model for a quantum Hall system \cite{QH}.
For $b=0$ it is natural to keep the cutoff $\Lambda$ finite: it has a
natural interpretation as a spatial cutoff, and $\nu= N/\Lambda$ is
the fermion density (`filling factor') of the QH system. The energy
eigenvalues are sums of two terms: the kinetic energy $\cE_0 =
|B|\sum_{j=1}^N (m_j-1)$ and the correlation energy $\cE_{\rm
corr}=2g\vv_{[\lambda]}$. Let $N<\nu$. Then obviously $\cE_0=0$ is
minimal if all $m_j=1$ (i.e.\ all fermions are in the lowest Landau
level).  Then all $\ell_j$ need to be different, and (for fixed
$\ell_j$'s) there is a unique eigenstate $|N\rangle$ with correlation
energy $\cE_{\rm corr} = gN(N-1)$. This obviously leads to a minimal
energy $\cE= gN(N-1)$ if $g\leq 0$, but for positive $g$ it can be
preferable have some fermions with $m_j> 1$: this increases the
kinetic energy but allows to decrease the correlation energy. For
example, if $m_j=j$ (all different: we put each electron in another
Landau level) and $\ell_j=\ell$ (all the same, to have a simple
specific example), $|N\rangle$ is eigenstate with total energy
$\cE'=N(N-1)|B|/2 - gN(N-1)$.  Obviously, $\cE'$ will be lower than
$\cE$ for sufficiently large $g$. Thus, if one increases the coupling
from $g<0$ zero to $g > |B|/4$ there must be some drastic change of
the ground state in between --- possibly a phase transition?  This QH
model becomes more interesting if one adds Hartree-Fock terms as in
Eq.\ \Ref{HF}: this allows to lift the degeneracy (fermions in the
same Landau level can repel each other, e.g.).

\item {\it Meaning of dimension?} From an abstract point of view, the
model we solved looks the same in all dimensions $2n+1$. This might
seem somewhat puzzling, in particular if one recalls that
renormalizability of quantum field theory models usually very much
depend on dimensions. This does not seem to be the case here: these
models therefore seem to challenge our usual expectations (and
prejudices) about locality and dimension in quantum field theory. I
should add that this is less unusual from a solid-state point of view:
the above-mentioned Hartree-Fock models also look the same in all
dimensions.

\end{itemize} 

\bigskip

\noindent {\bf Acknowledgments.} I would like to thank H.\ Grosse, J.\
Hoppe, J.\ Mickelsson, R.\ Szabo, T.\ Turgut, R.\ Wulkenhaar, and K.\
Zarembo for useful discussions. I am grateful to J.\ Hoppe for
comments on the manuscript and to J.\ Mickelsson for helping me with
group theory.  I thank R.\ Szabo for helpful suggestions about the
literature. I also acknowledge discussions with G.\ Semenoff quite
some time ago: he suggested to look for new fermion models which can
be solved using group theory. This work was supported in part by the
Swedish Science Research Council (VR) and the G\"oran Gustafssons
Foundation.

\bigskip

\noindent {\bf Appendix A. The star product of the Landau
eigenfunctions.} Fact~1 and Fact~2 in the main text are known since a
long time. However, due to the weight they carry in my discussion I
felt I should also include an (elementary) proof.  This is the purpose of
this Appendix.

We assume $2n=2$ and write $H_B\equiv H_1=P_1^2+P_2^2$ and
$H_{-B}\equiv H_2=\tilde P_1^2 + \tilde P_2^2$ where
\eqa
P_1=-i\partial_1 -|B| x_2,\quad P_2=-i\partial_2 +|B| x_1 \nonu
\tilde P_1=-i\partial_1 +|B| x_2,\quad \tilde P_2=-i\partial_2 -|B| x_1 
\eqaend
with $|B|>0$, $\partial_\mu=\frac{\partial}{\partial x^\mu}$ and
$(x^\mu)=(x^1,x^2)$ coordinates on $\R^2$.
We observe that 
\eqa
p_1 = \frac{1}{\sqrt{2|B|}}P_1,\quad q_1 = \frac{1}{\sqrt{2|B|}}P_2,\nonu
p_2 = \frac{1}{\sqrt{2|B|}}\tilde P_2,\quad q_2 = \frac{1}{\sqrt{2|B|}}\tilde P_1
\eqaend
give a representation of the Heisenberg algebra
${[}p_\mu,q_\nu{]}=-i\delta_{\mu,\nu}$ etc.  Thus
$H_1=2|B|(p_1^2+q_1^2)$ is just a harmonic oscillator Hamiltonian
which (on the Hilbert space of functions in two variables) is highly
degenerate. The operator $H_2=\tilde P_1^2+\tilde
P_2^2=2|B|(p_2^2+q_2^2)$ is the `complimentary harmonic
oscillator'allowing us to resolve this degeneracy. The Landau
eigenfunctions are the common eigenfunctions of $H_1$ and $H_2$. To
construct them we introduce creation-and annihilation operators
$a^\pm_\mu = \frac{1}{\sqrt2} (\mp ip_\mu+q_\mu)$ obeying the usual
commutation relations. Then $H_\mu=4|B|(a_\mu^+a^-_\mu +
\frac12)$. The common eigenfunctions of $H_1$ and $H_2$ are
therefore\footnote{The phase factor $(-i)^{m-1}$ is inserted for
convenience. I find it also convenient to label these states by
positive integers (i.e.\ what I call $\ell-1$ is usually called
$\ell$).}
\eq
|\ell,m\rangle \, = \, (-i)^{m-1} \frac{(a_1^+)^{\ell-1}}{\sqrt{(\ell-1)!}}\frac{(a_2^+)^{m-1}}{\sqrt{(m-1)!}}|0\rangle 
\eqend
where $a^-_\mu|0\rangle=0$ and $\ell,m$ positive integers.  The
eigenvalues are $4|B|(\ell-\frac12)$ and $4|B|(m-\frac12)$,
respectively.

We now compute the normalized Landau eigenfunctions in position space,
\eq
\phi_{\ell m}(\vx) =\, \langle\vx|\ell,m\rangle  
\eqend
where $\vx=(x^1,x^2)$. It is convenient to define
\eq
z= x_1+ix_2,\quad \bar z= x_1-ix_2  
\eqend
and to introduce the generating function
\eqa
F_{s,t}(\vx) = \sum_{\ell,m=0}^\infty \frac{s^\ell t^m}{\sqrt{\ell!m!}}\phi_{\ell+1, m+1}(\vx) = 
\nonu
= \sum_{\ell,m=0}^\infty \frac{s^\ell (-it)^m}{\ell!m!}\langle\vx|(a_1^*)^\ell (a_2^*)^m|0\rangle = 
\langle z,\bar z| e^{s a_1^* -it a_2^*}|0\rangle . 
\eqaend
Since $z=(a_1^* +ia_2)/\sqrt{|B|}$ and $\bar z=(a_1 -ia_2^*)/\sqrt{|B|}$ one gets 
\nonueqa
F_{s,t}(\vx) = \langle\vx| e^{s a_1^* -it a_2^*} e^{ t a_1 + i s a_2}|0\rangle 
=  \langle\vx|  e^{\sqrt{|B|}(sz + t\bar z)} e^{[s a_1^* -it a_2^*, t a_1 + i s a_2]/2}|0\rangle 
= \nonu =  e^{\sqrt{|B|}(sz + t\bar z)} e^{-st } \phi_{1,1}(\vx).  
\nonueqaend
The normalized ground state wave function $\phi_{1,1}$ can be computed
by solving $a_\mu\phi_{1,1}=0$ with $a_1= (\partial_1 -i\partial_2
+|B|(x_1-ix_2))/2\sqrt{|B|}$ and similarly for $a_2$. One thus finds
\eq F_{s,t}(\vx) = \sqrt{\frac{|B|}{\pi}} e^{-st}\, e^{\sqrt{|B|}(sz +
t\bar z)-|B||z|^2/2} \: .  \eqend
Using this generating function it is easy to check that the Landau
eigenfunctions thus defined are a complete orthonormal basis.

To prove Fact~1 we compute $F_{s_1,t_1}\star F_{s_2,t_2}$ assuming
$\theta=B^{-1}$. We use 
\eq
\label{star1}
(f\star g)(\vx) = 
(2\pi)^{-2} \int_{\R^2} d^2\vk \int_{\R^2} d^2\vq\, \hat f(\vk) \hat g(\vq) \, 
e^{ i |B|^{-1} (k_1 q_2-k_2 q_2) }\, 
e^{ i(\vk+\vq)\cdot\vx}   
\eqend
where $\hat f$ is the Fourier transform of $f$. We first compute ($K = k_1+ik_2$) 
\nonueqa
\hat F_{s,t}(\vk) = \int_{\R^2} \frac{d^2\vx}{2\pi}\, e^{-i\vk\cdot \vx}\, F_{s,t}(\vx) = 
\sqrt{\frac{1}{\pi |B|}} e^{st} \, e^{ i(sK + t\bar K)/\sqrt{|B|} - |K|^2/2|B| } 
\nonueqaend
(we computed a Gaussian integral).  Thus
\nonueqa
F_{s_1,t_1}\star F_{s_2,t_2} (\vx) = 
(2\pi)^{-2} \int_{\R^2} d^2\vk \int_{\R^2} d^2\vq\, e^{-i \theta(k_1 q_2-k_2 q_2) -i(\vk+\vq)\cdot\vx}\, \nonu
\times 
\frac{1}{\pi |B|}e^{s_1t_1 +s_2t_2} \,e^{ i(s_1K + t_1\bar K + s_2Q +t_2 \bar Q )/\sqrt{|B|} - 
(|K|^2+|Q|^2) /2|B| }  = \nonu = \frac{|B|}{ 2 \pi} \, e^{-s_1t_2+ s_2t_1}\, e^{ ( s_1 z + t_2\bar z) \sqrt{|B|} - 
|B| |z|^2/2  }  
\nonueqaend
(again a Gaussian integral), i.e.,
\eq
F_{s_1,t_1}\star F_{s_2,t_2} = \sqrt{\frac {|B|}{ 4 \pi} }  \, e^{s_2 t_1} \, F_{s_1,t_2}(\vx)  
\eqend
equivalent to 
\eq
\phi_{\ell_1,m_1}\star \phi_{\ell_2,m_2} =  \sqrt{\frac{|B|}{ 4 \pi} } \delta_{m_1,\ell_2}
\phi_{\ell_1,m_2} \, . 
\eqend
and completing our proof of Fact~1, including the normalization. Our
discussion in the beginning of this Appendix also provides a proof of
Fact~2.

\bigskip

\noindent {\bf Appendix B. Pedestrian solution of Eq.\ \Ref{ev}.}  In
the main text we give the general result for all $N$. To appreciate
this result and have a few special cases, I present here a brute-force
solution for small $N$.

One can represent $C_N$ \Ref{CN} by a $N!\times N!$-matrix in the
following way. Define a Hilbert space isomorphic to $\R^{N!}$ by
introducing the orthonormal basis $|P)$, $P\in S_N$. Then
$\pi_{P}(P',P'') := (P',P\cdot P)$ defines a representation $P\to
\pi_P$ by $N!\times N!$ matrices (the reader familiar with group
theory will recognize that $\pi$ is just the regular representation of
$\pi$). We thus can solve \Ref{ev} by constructing and diagonalizing
the $N!\times N!$-matrix $C_N:= \sum_{1\leq j<k\leq N}\pi_{(jk)}$. We
now describe the results for $N\leq 4$ and give a few examples for
eigenfunctions.

For $N=0$ and $N=1$ it is easy to find all eigenstates and eigenvalues
of $\cH$,
\eq \cH\Omega= 0 \qquad (N=0) \eqend
and 
\eq \cH
A^\dag_{\ell m}\Omega = (E_\ell +E_m)A^\dag_{\ell m}\Omega \qquad
(N=1).  \eqend 
The first non-trivial case is $N=2$. Labeling the basis
$|P)$, $P\in S_2$, in the following way,
$|12)\equiv \hat 1$ and $|21)\equiv \hat 2$, 
(we use an obvious short-hand notation for permutations on the l.h.s.\
of these equations) it is easy to see that
$$
C_2 = \left(\bma{cc} 0&1\\1&0\ema \right) ,  
$$
which has eigenvectors $(a_{\hat 1},a_{\hat 2})=(1,\pm 1)$ corresponding to the eigenvalues 
\eq
\vv=\pm 1\quad (N=2) . 
\eqend
From that one obtains the following eigenstates of $\cH$,
\eq
\label{zpm} 
|2)_\pm = (a_{\hat 1} I + a_{\hat 2} T_{(12)} ) |2\rangle =
\left( A^\dag_{\ell_1 m_1} A^\dag_{\ell_2 m_2} \pm
A^\dag_{\ell_1 m_2} A^\dag_{\ell_2 m_1}\right) \Omega \eqend
(we identified $T_{\hat 1} \equiv T_{12}=I$ and $T_{\hat
2}=T_{21}=T_{(12)}$) corresponding to the eigenvalues $\cE=\cE_0\pm
2g$. Note that if $m_1=m_2$ then $|N\rangle_+ =0$ and if
$\ell_1=\ell_2$ then $|N\rangle_- =0$ (due to the Pauli
principle). Thus there are two independent eigenstates only if
$m_1\neq m_2$ and $\ell_1\neq \ell_2$, otherwise there is only
one. Still, the $N=2$-eigenstates thus constructed provide a complete
orthonormal basis in the $N=2$-subspace of our fermion Hilbert space.

For $N=3$ we label the basis as $|123)=\hat 1$, $|312)=\hat 2$,
$|231)=\hat 3$, $|213)=\hat 4$, $|132)=\hat 5$, and $|321)=\hat 6$,
and we obtain
$$
C_3 = {\tiny \left(\bma{cccccc} & & & 1&1&1\\ & & & 1&1&1\\ & & & 1&1&1\\  
1&1&1  & & &\\ 1&1&1  & & &\\ 1&1&1  & & &
 \ema \right) }  
$$
(the matrix elements not written are zero).  The eigenvalues of this
matrix are
\eq 
\vv= 3(1),0(4),-3(1) \qquad (N=3)
\eqend 
with the numbers in the parenthesis indicating the
multiplicities.  It is not difficult to also write down the
corresponding eigenstates. For example, $(a_{\hat 1},\ldots,a_{\hat
6})=(1,-1,0,0,0,0)$ is an eigenvector of $C_3$ with eigenvalue $\vv=0$,
and the corresponding eigenstate of $\cH$ with eigenvalue $\cE=\cE_0$ is
\eq
\left(T_{123} -T_{312}\right)|3\rangle = \left( A^\dag_{\ell_1 m_1}
A^\dag_{\ell_2 m_2}A^\dag_{\ell_3 m_3} \pm A^\dag_{\ell_1 m_3}
A^\dag_{\ell_2 m_1}A^\dag_{\ell_3 m_2} \right) \Omega 
\eqend
etc. If all three $m_j$'s are different and also the three $\ell_j$'s,
these eigenstates all are linearly independent, but otherwise the
number of linearly eigenvectors can be less than 6 and not all
eigenvalues $\cE_0+2g\vv$ are realized.  I also constructed and
diagonalized the $24\times 24$-matrix $C_4$ and found the following
eigenvalues (and multiplicities),
\eq 
\vv= 6(1),2(9),0(4),-2(9),-6(1) \qquad (N=4) .  
\eqend 
Again, the corresponding 24 eigenstates are linearly independent only
if the four $m_j$'s are all different and the same for the $\ell_j$'s.
At this point this brute-force approach clearly becomes too
cumbersome. Fortunately for us, the eigenvalues and eigenvectors of
all the matrices $C_N$ are all known from group theory.

\end{document}